\documentclass[sigconf]{acmart}

\AtBeginDocument{%
  \providecommand\BibTeX{{%
    \normalfont B\kern-0.5em{\scshape i\kern-0.25em b}\kern-0.8em\TeX}}}

\makeatletter
\renewcommand\@copyrightpermission{}
\makeatother
\setcopyright{none}

\usepackage[ruled, linesnumbered]{algorithm2e}
\usepackage{caption,url,hyperref}
\usepackage{enumitem}
\usepackage{graphicx}
\usepackage{xcolor} 
\usepackage{hyperref}
\hypersetup{
    colorlinks=true,
    linkcolor=red,
    urlcolor=red,
}

\begin{document}
\title{ConCLVD: Controllable Chinese Landscape Video Generation via Diffusion Model}


\author{Ding-Ming Liu}
\affiliation{%
  \institution{Department of Artificial Intelligence\\Xiamen University}
  \country{P. R. China}}

\author{Shao-Wei Li}
\affiliation{%
  \institution{Department of Artificial Intelligence\\Xiamen University}
  \country{P. R. China}}

\author{Ruo-Yan Zhou}
\affiliation{%
  \institution{Department of Computer Science\\Xiamen University}
  \country{P. R. China}}

\author{Li-Li Liang}
\affiliation{%
  \institution{Department of Computer Science\\Xiamen University}
  \country{P. R. China}}

\author{Yong-Guan Hong}
\affiliation{%
  \institution{Department of Artificial Intelligence\\Xiamen University}
  \country{P. R. China}}

\author{Fei Chao}
\affiliation{%
  \institution{Department of Artificial Intelligence\\Xiamen University}
  \country{P. R. China}}

\author{Rongrong Ji}
\affiliation{%
  \institution{Department of Artificial Intelligence\\Xiamen University}
  \country{P. R. China}}

\renewcommand{\shortauthors}{}

\begin{abstract}

Chinese landscape painting is a gem of Chinese cultural and artistic heritage that showcases the splendor of nature through the deep observations and imaginations of its painters. Limited by traditional techniques, these artworks were confined to static imagery in ancient times, leaving the dynamism of landscapes and the subtleties of artistic sentiment to the viewer's imagination. Recently, emerging text-to-video (T2V) diffusion methods have shown significant promise in video generation, providing hope for the creation of dynamic Chinese landscape paintings. However, challenges such as the lack of specific datasets, the intricacy of artistic styles, and the creation of extensive, high-quality videos pose difficulties for these models in generating Chinese landscape painting videos. In this paper, we propose CLV-HD (Chinese Landscape Video-High Definition), a novel T2V dataset for Chinese landscape painting videos, and ConCLVD (Controllable Chinese Landscape Video Diffusion), a T2V model   that utilizes Stable Diffusion. Specifically, we present a motion module featuring a dual attention mechanism to capture the dynamic transformations of landscape imageries, alongside a noise adapter to leverage unsupervised contrastive learning in the latent space. Following the generation of keyframes, we employ optical flow for frame interpolation to enhance video smoothness. Our method not only retains the essence of the landscape painting imageries but also achieves dynamic transitions, significantly advancing the field of artistic video generation. The source code and dataset will be released soon.

\vspace{-1.0em}
\end{abstract}





\begin{teaserfigure}
  \begin{center}
    \includegraphics[width=0.85\linewidth]{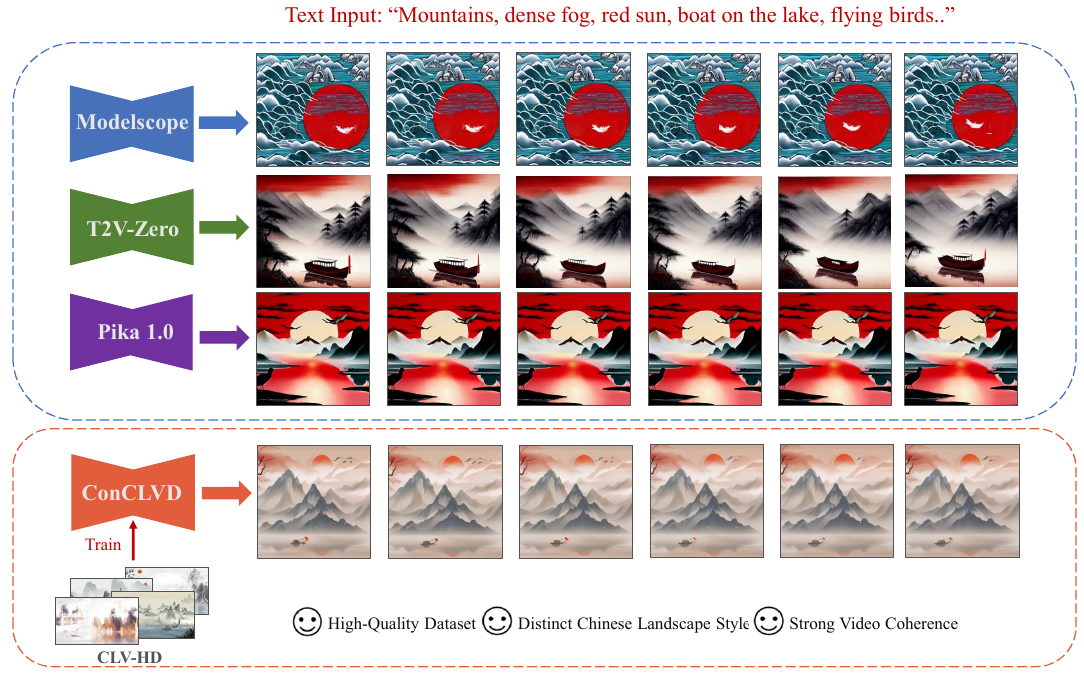}
    \caption{Chinese Landscape Video Generation.}
    \Description{.}
    \label{fig:top}
  \end{center}
\end{teaserfigure}


\maketitle

\section{Introduction}

Chinese landscape painting is a vital component of China's cultural legacy. It stands out for blending philosophical depth with aesthetic appeal, employing distinctive light and shadow techniques, and mastering sophisticated ink methods. However, as modern technology evolves and people's aesthetic preferences become more diverse and modernized, the inheritance of this traditional art form faces challenges. Recently, advancements~\cite{ChangMM2022,Xue2020EndtoEndCL,Lin2018TransformAS,Wang2019ChinaStyleAM,Wang2023CCLAPCC,LiLZWXWZP0J23} in deep learning technology offer new possibilities for the modern inheritance of this ancient art form through innovative image generation of Chinese landscape paintings.

However, the charm of landscape painting lies not only in its static beauty but also in the dynamic imagination and profound sense of the picture it inspires. By utilizing the dynamic medium of video, we integrate the traditional charm of Chinese landscape painting with the innovative power of modern technology, infusing new life into the art form and enabling it to exhibit a richer and more distinct layering of beauty in motion. This medium not only attracts young audiences and promotes the dissemination and advancement of art, but also opens new avenues of expression and inspiration, offering valuable perspectives and tools for cultural heritage, education, and artistic innovation. Therefore, developing high-quality dynamic Chinese landscape painting video generation is particularly important.

Although Text-to-Video (T2V) ~\cite{An2023LatentShiftLD,Singer2022MakeAVideoTG,Wang2023VideoFactorySA,Xing2023MakeYourVideoCV,Zhou2022MagicVideoEV,Deng2023EfficiencyoptimizedVD,Deng2023MVDiffusionMV} is widely studied in various fields, it still falls short in capturing the essence of Chinese landscape painting videos. First, the absence of specific datasets for landscape painting styles hampers the accurate depiction of their unique style and artistic traits. Second, there is a disparity in balancing the local and global coherence in videos, which is crucial for depicting landscapes' natural transitions. Lastly, current methods fall short in capturing the similarities and differences between frames, which affects the fluidity and authenticity of the videos, ultimately impacting their artistic appeal.

To tackle the specific challenges of generating Chinese landscape painting videos, notably the scarcity of datasets, we develop CLV-HD (Chinese Landscape Video-High Definition), a dedicated text-to-video dataset with around 1,300 curated entries. This dataset aims to fill the technological void in emulating Chinese landscape painting's unique aesthetics, offering a vast resource to enhance deep learning models' ability to replicate this art form's distinct style. CLV-HD encompasses a range from traditional to modern styles and details the videos' dynamic elements, supporting our tech exploration and providing new tools for the art's modern evolution and innovation.

Technologically, drawing inspiration from AnimateDiff~\cite{Guo2023AnimateDiffAY}, we merge a bespoke motion module with frozen Stable Diffusion, blending Versatile Attention~\cite{Vaswani2017AttentionIA,Guo2023AnimateDiffAY}  and Sparse-Causal Attention~\cite{Child2019GeneratingLS,Hertz2022PrompttoPromptIE,Ma2023FollowYP} for nuanced temporal data analysis. Versatile Attention assesses the entire time series for global patterns, while Sparse-Causal Attention hones in on local causality, enhancing future event predictions. This synergy improves global and local trend analysis accuracy. To ensure the similarity of adjacent frames and the dissimilarity of distant frames, we employ a noise contrastive learning approach in latent space, using a noise adapter to construct contrastive learning samples based on the predicted noise output from the de-noising U-Net. In addition, we introduce a frame interpolation strategy based on optical flow projection for smoother frame transitions. Experiments indicate that our ConCLVD framework is capable of producing high-quality, varied, and highly coherent videos that embody the style of Chinese landscape painting.

Overall, the main contributions of this paper are as follows:
\begin{itemize}
    \item We innovatively address a new task of generating videos in the style of Chinese landscape painting from text by designing our framework ConCLVD, which combines a motion module based on Stable Diffusion and a contrastive learning strategy in latent space, realizing cost-effective and efficient artistic video generation.
    \item We have created a new high-quality dataset named CLV-HD, which contains about 1,300 videos of Chinese landscape paintings, with each video accompanied by detailed text descriptions.
    \item We propose a new interpolation method using sparse optical flow projection, a plug-and-play approach that doesn't require training, which effectively enhances video smoothness by filling in intermediate frames accurately and efficiently.
    \item Our model requires lower computational power, allowing us to offer a more affordable implementation and experience for more people interested in generating Chinese landscape painting videos.

\end{itemize}

\section{Related Work}




\subsection{Text-to-Video Diffusion Model}

Despite significant progress in Text-to-Image (T2I) generation technology, Text-to-Video (T2V) generation technology remains relatively lagging. This lag is mainly due to the lack of large-scale and high-quality text-video pairing datasets, as well as the high-dimensional modeling complexity of video data. Early T2V research~\cite{Mittal2016SyncDRAWAV,Pan2017ToCW,Marwah2017AttentiveSV,Gupta2018ImagineTS,Liu2019CrossModalDL} primarily focused on generating simple video content, but as time progressed, researchers began exploring more advanced techniques to address the challenges of T2V generation.

Recently, several studies~\cite{Geyer2023TokenFlowCD,Qi2023FateZeroFA,Wu2023LAMPLA} have shifted towards using the knowledge of pre-trained T2I models, simplifying the construction of T2V models by performing the diffusion process in the latent space. In particular, T2V diffusion models have adopted a spatio-temporal separable architecture, inheriting the spatial operations of pre-trained T2I models and reducing the complexity of constructing intricate models. Models such as AnimateDiff~\cite{Guo2023AnimateDiffAY} have enhanced the capability of generating dynamic videos by introducing innovative motion modeling modules, and new control mechanisms have been introduced to increase the application flexibility and quality of generation. The development of these diffusion models marks significant progress in T2V generation technology. Although they still require extensive training and face flexibility limitations in the application, they offer valuable perspectives and methods for addressing core challenges in T2V generation, propelling further exploration and innovation in the field.

\subsection{Chinese Landscape Generation}

GAN-based methods for generating images of Chinese landscape paintings have been extensively studied~\cite{Xue2020EndtoEndCL,Lin2018TransformAS,Wang2019ChinaStyleAM,Yuan2022LearningTG,Lin2018TransformAS,Zhou2019AnIA}. For example, Xue et al.~\cite{Xue2020EndtoEndCL} uses GANs to generate landscape paintings in two stages: initially generating basic outline sketches of landscape paintings from random noise; then transforming these sketches into final paintings through edge-to-painting conversion. Polaca~\cite{Yuan2022LearningTG} is a poetry-oriented landscape painting generation model, transforming poetic texts into landscape images using GANs, generating matching calligraphy, and finally merging both to form complete art pieces. In recent years, diffusion models have achieved commendable results in generating images of Chinese landscape paintings. CCLAP~\cite{Wang2023CCLAPCC} uses texts and reference images as conditions, employing latent diffusion models to generate Chinese landscape paintings with controllable content and style. However, to date, the generation of Chinese landscape painting videos has received limited attention. To our knowledge, we are the first to utilize diffusion models for generating videos of Chinese landscape paintings. Furthermore, unlike previous work where images serve as the primary guiding condition, our focus is on capturing the transformation of scenery, transitions of light and shadow, and other dynamic elements in landscape paintings. Hence we frame the task as a T2V generation task.

\section{Method}

\subsection{Preliminary}
Stable Diffusion, based on the Latent Diffusion Model (LDM)~\cite{latentdiffusionmodel}, is a type of diffusion model that enhances image generation. The process starts by encoding an input image $x_0$ into a latent space $z_0 = E(x_0)$ using an encoder $E$. This latent representation $z_0$ is then modified through a series of steps according to the equation:
\begin{equation}\label{eq_latent_representation}
q(z_t | z_{t-1}) = \mathcal{N}(z_t; \sqrt{1-\beta_t} z_{t-1}, \beta_t I),
\end{equation}
where $z_t$ represents the latent state at step $t$, and $\beta_t$ are hyperparameters controlling the noise added at each step. The process can be summarized as:
\begin{equation}\label{eq_process}
z_t = \sqrt{\overline{\alpha_t}} z_0 + \sqrt{1 - \overline{\alpha_t}} \epsilon, \epsilon \sim \mathcal{N}(0, I),
\end{equation}
where $\overline{\alpha_t} = \prod_{i=1}^{t} \alpha_i$ and $\alpha_t = 1 - \beta_t$.

Stable Diffusion adopts the vanilla training objective as proposed in DDPM~\cite{dhariwal2021diffusion}, which is expressed as:
\begin{equation}\label{eq_sd_training}
\mathcal{L} = \mathbb{E}_{E(x_0), y, \epsilon \sim \mathcal{N}(0, I), t} \left[ \|\epsilon - \epsilon_{\theta}(z_t, t, \tau_{\theta}(y))\|^2 \right] ,
\end{equation}
where $y$ is the text description and $\tau_{\theta}(\cdot)$ is a text encoder based on the CLIP~\cite{CLIP} ViT-L/14 model. The architecture features a UNet~\cite{unet} with downscaling and upscaling blocks, incorporating 2D convolutional, self-attention, and cross-attention layers.

DDIM~\cite{Song2020DenoisingDI} is proposed to accelerate the sampling process by optimizing the noise reduction steps. It transforms a random noise vector \( z_T \) into a coherent latent representation \( z_0 \) through a defined sequence of timesteps \( t: T \rightarrow 1 \), which is defined as:
\begin{equation}\label{ddim}
z_{t-1}=\sqrt{\alpha_{t-1}} \frac{z_{t}-\sqrt{1-\alpha_{t}} \epsilon_{\theta}(z_t, t, \tau_{\theta}(y))}{\sqrt{\alpha_{t}}}+\sqrt{1-\alpha_{t-1}} \epsilon_{\theta}(z_t, t, \tau_{\theta}(y)) ,
\end{equation}
where $\alpha_{t}$ is a parameter for noise scheduling, $\epsilon_{\theta}(z_t, t, \tau_{\theta}(y))$ is the predicted noise within the networks’ latent space.

AnimateDiff~\cite{Guo2023AnimateDiffAY} extends the Stable Diffusion model for T2V tasks by integrating a motion modeling module for video data handling. It progresses from processing four-dimensional image batches to handling five-dimensional video tensors $(batch \times channels \times frames \times height \times width)$. The transformation adapts each 2D convolution and attention layer from the base image model into pseudo-3D layers that focus solely on spatial aspects. This adaptation involves reorganizing the frame dimension to merge with the batch dimension, which permits independent frame analysis. Subsequently, feature maps undergo a transformation to a $(batch \times height \times width) \times frames \times channels$ configuration. This step sets the stage for the motion module, aimed at ensuring consistent motion and content stability across frames.



\begin{figure*}[!ht]
    \centering
    \includegraphics[width=.8\linewidth]{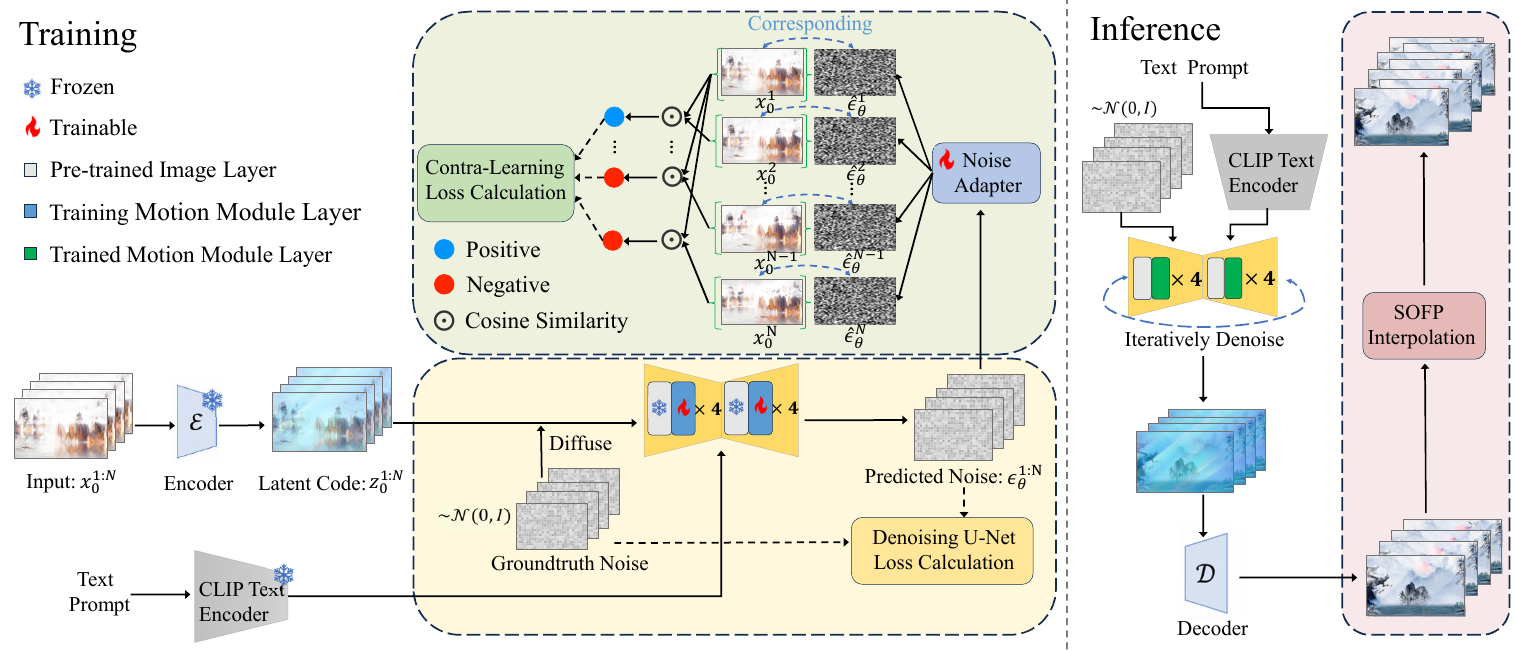}
    \captionsetup{font=small}
    \caption{Overview of ConCLVD. Left: the architecture. ConCLVD integrates a trainable motion module based on a frozen Stable Diffusion and introduces a noise adapter to accommodate contrast learning of noise in latent space. Right: the inference framework. The video is generated by the denoising U-Net integrated with the motion module and enhanced by sparse optical flow projection (SOFP) interpolation technology.}\label{fig_atten}
\end{figure*}

Motion modules are embedded throughout the U-shaped network, using vanilla temporal transformers with self-attention blocks operating along the temporal axis. The training objective is to minimize the following loss function:
\begin{equation}\label{eq_loss_function}
\mathcal{L} = \mathbb{E}_{E(x_0^{1:N}), y, \epsilon \sim \mathcal{N}(0, I), t} \left[ \| \epsilon - \epsilon_{\theta}(z_t^{1:N}, t, \tau_{\theta}(y)) \|_2^2 \right] ,
\end{equation}
where $x_0^{1:N}$ represents a sample of video data, and $z_t^{1:N}$ is the latent code obtained by adding noise to the initial latent code $z_0^{1:N}$ at time step $t$. During training, the pre-trained weights of the base T2I model are frozen to maintain consistency in the feature space.


\subsection{CLV-HD: Text-to-Chinese Landscape Painting Video Dataset}

Diverse text-video datasets are essential for developing high-quality T2V generation models. However, there is a notable shortage of such datasets for Chinese landscape painting videos, limiting the fusion of traditional Chinese art with modern technology. The lack of comprehensive datasets with textual annotations hampers the creation of high-quality Chinese landscape painting videos.

Our work fills this notable gap by introducing CLV-HD, a groundbreaking text-to-Chinese landscape painting video dataset, now publicly available. It includes 1,300 text-video pairs, sourced from open domains in high-definition. We collected 210 high-resolution clips from Chinese water ink animations and over 400 original videos from YouTube. To manage complex scene transitions, we meticulously segmented these into 1,300 single-scene clips, enhancing training utility. Acknowledging the videos' intricate details, we manually annotate each text to accurately match the video content.

Our CLV-HD dataset effectively addresses the scarcity of specialized datasets for generating Chinese landscape painting videos from textual descriptions. This pioneering effort fosters the convergence of Chinese traditional art and artificial intelligence, catalyzing new avenues for scholarly inquiry and practical application.

\subsection{Controllable Chinese Landscape Video Generation}


Our model integrates two key components: a denoising U-Net and a Noise-Adapter. The denoising U-Net uses pre-trained Stable Diffusion weights for enhanced image layer processing. To capture motion, a motion module layer follows each image layer, extending the T2I functionality to T2V and enabling dynamic motion synthesis. Our training incorporates unsupervised contrastive learning through the Noise-Adapter, which refines the noise output from the U-Net into contrastive learning samples, enhancing noise representation in the latent space. The structure of the motion module and the contrastive learning strategy are detailed in Section~\ref{sec_motiontraining}.

During the training process, we keep the image layers of the denoising U-Net frozen and only update the motion module layers and the Noise-Adapter. During the inference phase, the model creates coherent landscape painting animations through an iterative denoising process. The video's length and smoothness are then further refined using a projection frame interpolation technique based on sparse optical flow, as detailed in Section~\ref{sec_projection}. This procedure not only amplifies the visual appeal of the animation but also enhances the video's coherence and aesthetic value, providing a novel method for dynamically presenting landscape paintings.

\subsection{Motion Module and Training}
\label{sec_motiontraining}
\subsubsection{\textbf{Motion Module Design}}

\begin{figure}[!t]
    \centering
    \includegraphics[width=1\linewidth]{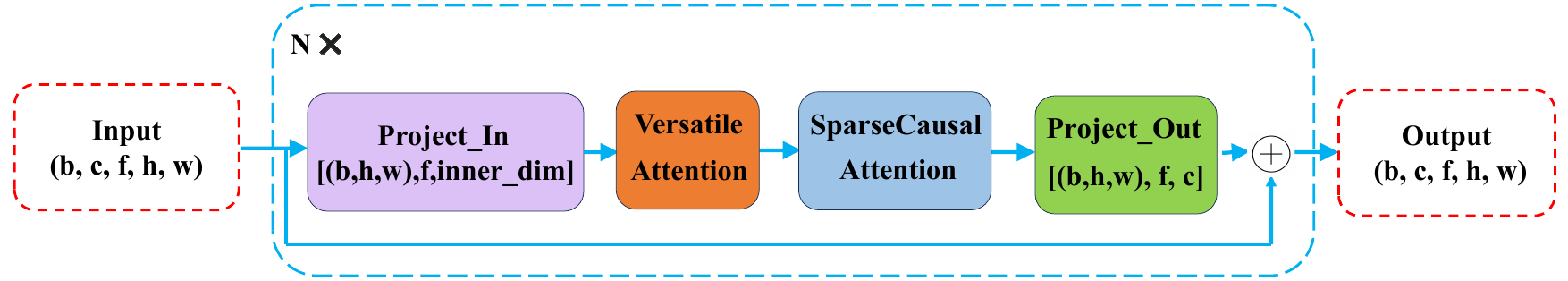}
    \captionsetup{font=small}
    \caption{Design of Motion Module. The motion module is inserted following each image layer of the pre-trained SD to process video data.}
    \label{module}
    \vspace{-1.5em}
\end{figure}

In our motion module, the core design principle is to capture dynamic changes between frames through effective information exchange. We utilize standard temporal transformers as the core architecture, to facilitate efficient information exchange between frames to capture the dynamic information that evolves within video sequences. To achieve this, we integrate two key attention mechanisms: Versatile Attention~\cite{Vaswani2017AttentionIA,Guo2023AnimateDiffAY} and Sparse-Causal Attention~\cite{Child2019GeneratingLS,Hertz2022PrompttoPromptIE,Ma2023FollowYP} , which are specifically optimized for time series data and analyze the temporal dimension by processing feature tensors in a particular shape configuration. The structure of the motion module is demonstrated in Figure.~\ref{module}.

The motion module is inserted between the image layers of the pre-trained Stable Diffusion. When a batch of data passes through our motion module, the video data undergoes reshaping and dimension transformation in Project\_In module, converting the input tensor from from $[batch, channels, frames, height, width]$ to $[(batch \times height \times width) \times frames \times {inner\_dim}]$. Through this approach, the model can meticulously analyze interactions between individual frames and capture local and global features using the designed attention mechanism. Finally, the Project\_Out module reverts the processed data back to its original dimensions and adds it to the initial input as the output employing a residual architecture.

The Versatile Attention~\cite{Vaswani2017AttentionIA,Guo2023AnimateDiffAY}  adapts the traditional self-attention framework for time-series data. The mechanism computes relationships between different time points, aiding the model in understanding video dynamics.

The Sparse-Causal Attention~\cite{Child2019GeneratingLS,Hertz2022PrompttoPromptIE,Ma2023FollowYP} mechanism focuses on simulating the causal relationships within the frame sequence, paying attention only to the previous frame preceding the current frame, thereby maintaining temporal causality and allowing the model to effectively understand and predict variations in the sequence without introducing future information.

The mathematical expressions for both attention mechanisms can be unified as follows:
\begin{equation}\label{eq_attation}
    z = \text{Attention}(Q,K,V) = \text{Softmax}\left(\frac{QK^T}{\sqrt{d}}\right) \cdot V,
\end{equation}
where $Q = W^Q z_{v_i}$, $K = W^K z_{v_j}$, $V = W^V z_{v_j}$, with $z_{v_i}$ and $z_{v_j}$ being the processed latent representations of the $i^{th}$ and $j^{th}$ frames, respectively. $W^Q$, $W^K$, and $W^V$ are learnable matrices that project the input into query, key, and value, and $d$ is the dimensionality of the key and query features. For Sparse-Causal Attention, a masking mechanism is applied to include only information from the previous frame that has a causal influence on the current frame.\\ The Versatile Attention mask is defined as:
\begin{equation}\label{m1}
\left(\mathrm{M}_{\text {Versatile }}\right)_{i j}=1.
\end{equation}
The Sparse-Causal Attention mask is defined as:
\begin{equation}\label{m2}
\left(\mathrm{M}_{\text {Sparse-Causal }}\right)_{i j}=\left\{\begin{array}{ll}1 & \text { if } j=i-1 \\0 & \text { otherwise }\end{array}\right . .
\end{equation}

\begin{figure}[!t]
    \centering
    \includegraphics[width=1\linewidth]{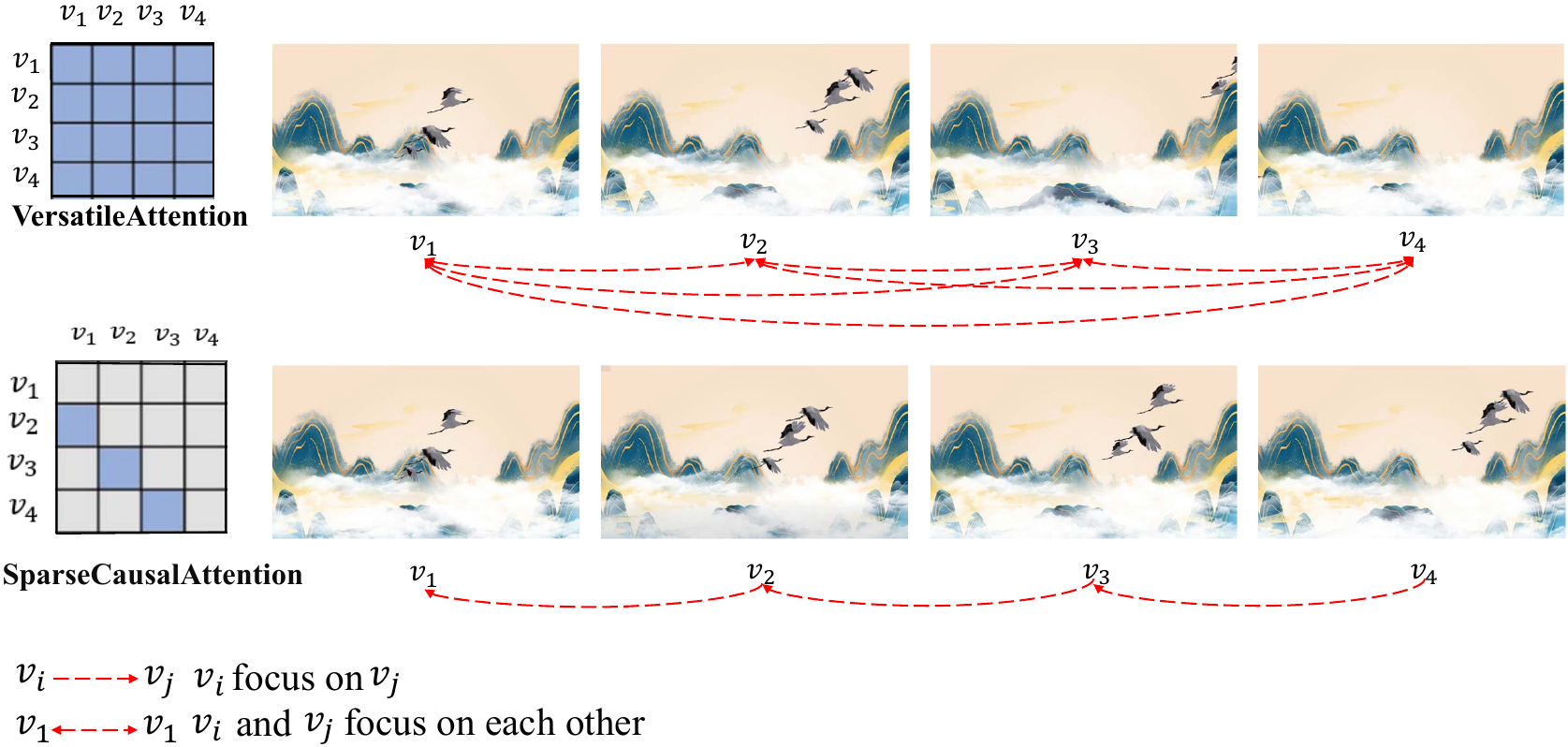}
    \captionsetup{font=small}
    \caption{Detailed explanation of the attention mechanism. The above represents Versatile Attention, where each frame is related to every other frame; the below represents Sparse-Causal Attention, where each frame only focuses on its previous frame.}
        \label{fig_attention}
    \vspace{-1.0em}
\end{figure}

Specifically, when we implement the Versatile Attention mechanism, we derive the query feature from frame $z_{v_i}$, and the key and value features from the first frame $z_{v_1}$ to the last frame $z_{v_T}$. When we implement the Sparse-Causal Attention mechanism, we derive the query feature from frame $z_{v_i}$, and the key and value features from the preceding frame $z_{v_{i-1}}$ (See Figure.~\ref{fig_attention} for a detailed visual depiction).

Through the combination of these two attention mechanisms, our motion modeling can not only capture the temporal dimensional features within video data but also ensure the consistency of content and smoothness of motion.

\subsubsection{\textbf{Unsupervised Contrastive Learning of Noise Corresponding to Video Frames in Latent Space}}
We introduce a contrastive learning strategy to deepen the model's grasp of dynamic transitions between video frames. We propose a unique approach for contrastive learning that leverages the UNet's distinct noise predictions for each frame. Specifically, we process the predicted noise vectors through a noise adapter, a method inspired by SimCLR\cite{simCLR} and SeCo\cite{SeCo}, to generate samples from these vectors. The essence of this strategy is to treat consecutive frames as positive pairs and frames where the frame number difference exceeds a threshold as negative pairs, using their respective noise vectors to create targeted training samples. It's important to mention that our exploration of different network designs for the noise adapter has led us to conclude that a basic MLP suffices. We leave the search for a better noise adapter to future works. Specifically, for the predicted noise $\epsilon_{\theta}^{i}$  of the $i^{th}$ frame output by the model, we assume that it passes through a noise adapter to obtain the corresponding contrastive learning sample $\hat{\epsilon}_{\theta}^{i}$. We calculate the similarity at step $t$ using the cosine similarity equation:
\begin{equation}
{r_t}^{(i,j)} = \frac{\hat{\epsilon_\theta}^i \cdot \hat{\epsilon_\theta}^j}{\|\hat{\epsilon_\theta}^i\| \|\hat{\epsilon_\theta}^j\|},
\end{equation}
where $\hat{\epsilon_\theta}^i$ and $\hat{\epsilon_\theta}^j$ are two vectors output by noise adapter for the $i^{th}$ frame and $j^{th}$ frame. Specifically, we calculate the contrastive learning loss for the noise vectors corresponding to all frames except the last one as anchors and calculate the average. For the noise vector corresponding to the $i^{th}$ frame, we have the following definition:
\begin{equation}\label{cl}
{l_t}^{(i)} =-\log \frac{\exp ({r_t}^{(i,i+1)} / \tau)}{\exp ({r_t}^{(i,i+1)} / \tau)+\sum_{k=1}^{N} 1_{\{|k-i|>m\}} \exp ({r_t}^{(i,k)} / \tau)},
\end{equation}
where $m$ represents the frame sequence threshold for negative samples, and $\tau$ denotes the temperature parameter for contrastive learning. The overall contrastive learning loss can be defined as:
\begin{equation}\label{conloss}
\mathcal{L}_{con}=\frac{1}{N-1} \sum_{k=1}^{N-1} {l_t}^{(k)},
\end{equation}
where $N$ is the total number of frames in the video. Through this contrastive learning of noise in the latent space, the model is encouraged to learn rich and robust temporal sequence feature representations without explicit annotations.

\subsubsection{\textbf{Training Objectives}}
In this section, we detail the training objectives of ConCLVD. We sample videos to obtain a sequence of frames $x_0^{1:N}$, and encode each frame into the latent space $z_0^{1:N}$ using a pre-trained Variational Autoencoder initially. Next, the latent codes are noised using the defined forward diffusion schedule: $\text z_t^{1:N} = \sqrt{\bar{\alpha}_t} z_0^{1:N} + \sqrt{1 - \bar{\alpha}_t} \epsilon $. During the training process, our network with the motion module receives noisy latent codes and corresponding textual prompts as input, predicting the noise intensity added to the latent codes. This process uses the L2 loss for calculation, and the loss for the motion module is defined as:
\begin{equation}\label{l1}
    \mathcal{L}_{diff} = \mathbb{E}_{\mathbb{E}(x_0^{1:N}),y,\epsilon \sim \mathcal{N}(0,I),t} \left[ \| \epsilon - \epsilon_{\theta}(z_t^{1:N}, t, \tau_{\theta}(y)) \|_2^2 \right].
\end{equation}
Note that during optimization, the pre-trained weights of the base Stable Diffusion are frozen. Only the motion module and noise adapter are trainable. Combining the diffusion model with the contrastive learning strategy, the overall objective function of the ConCLVD can be formulated as follows:
\begin{equation}\label{eq_overal_loss}
\mathcal{L} =\lambda_{diff} \cdot \mathcal{L}_{diff} +\lambda_{con} \cdot \mathcal{L}_{con}.
\end{equation}
The hyperparameters $\lambda_{diff}$ and $\lambda_{con}$ are used to balance different losses and achieve improved performance.

\subsection{Frame Interpolation Based on Sparse Optical Flow Projection}
\label{sec_projection}

\begin{figure}[!t]
    \centering
    \includegraphics[width=1\linewidth]{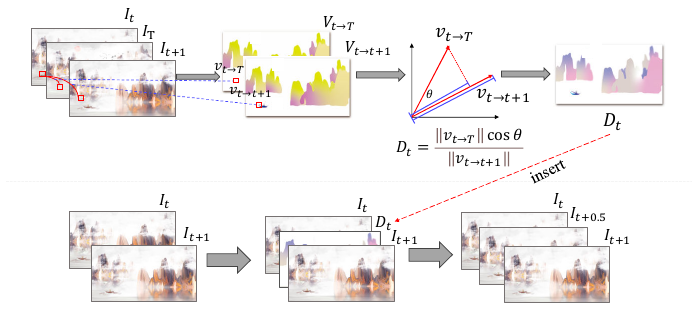}
    \captionsetup{font=small}
    \caption{Illustration of of SOFP Interpolation. $V_{(t \rightarrow T)}$ is the optical flow from $I_t$ to $I_T$ and $V_{(t \rightarrow t+1)}$ is the optical flow from $I_t$ to $I_{(t+1)}$. The projection result of $V_{(t \rightarrow T)}$ onto $V_{(t \rightarrow t+1)}$ is used to insert a new frame $I_{(t+0.5)}$ between two consecutive frames $I_t$ and $I_{(t+1)}$.
}
    \vspace{-1.4em}
\end{figure}

In the field of video processing, frame interpolation techniques based on optical flow~\cite{Park2021AsymmetricBM,Huang2020RealTimeIF,zhong2023clearer} are commonly used methods aimed at improving the visual experience. We propose a frame interpolation strategy that employs optical flow projection. This approach uses information from two adjacent frames along with the last frame of the video to interpolate additional frames between the two adjacent ones, aiming to achieve smooth transitions and enhance the continuity of the video playback.

Optical flow is defined as the motion vector of pixel points between two frames. For two images, the optical flow vector indicates the new position of each pixel in the first image within the second frame. Given a pair of consecutive frames $I_t$ and $I_{t+1}$, we calculate the optical flow $V_{t \rightarrow t+1}$ between the two frames. Similarly, we calculate the optical flow $V_{t \rightarrow T}$ between frame $I_t$ and frame $I_T$.

In our research, we employ the advanced optical flow estimation algorithm RAFT~\cite{RAFT} to precisely calculate the motion distance between pixels. Given a set of image triplets $\{I_t, I_T, I_{t+1}\}$, our primary task is to estimate two sets of optical flows: one from $I_t$ to $I_T$, denoted as $V_{t \rightarrow T}$, and the other from $I_t$ to $I_{t+1}$, denoted as $V_{t \rightarrow t+1}$. For each pixel point $(x, y)$ in the image, we project the motion vector $V_{t \rightarrow T}(x, y)$ from $I_t$ to $I_T$ onto the motion vector $V_{t \rightarrow t+1}(x, y)$ from $I_t$ to $I_{t+1}$. Thus, we define the distance mapping $D_t(x, y)$ as the ratio between these two vectors:
\begin{equation}\label{eq_distance}
D_t(x, y) = \frac{\|V_{t \rightarrow T}(x, y)\| \cos \theta}{\|V_{t \rightarrow t+1}(x, y)\|},
\end{equation}
where $\theta$ is the angle between these two vectors. The projection results are used for interpolation between frame $I_t$ and frame $I_{t+1}$.

The entire method of ConCLVD is summarized in Alogorithm~\ref{alg_1}.

\begin{algorithm}[!t]
\caption{ConCLVD's main learning algorithm}

\SetKwInput{Input}{Input} 
\Input{%
\\
   \textendash\ $z_{0}^{1:N}$: Latent code from source video \\
   \textendash\ $\tau_{\theta}(y)$: CLIP-encoded text prompt for the input video\\

}

\SetKwInput{Hyperparameters}{Hyperparameters} 
\Hyperparameters{%
\\
  \textendash\ $\tau$: Temperature parameter used in contrastive learning \\
   \textendash\ $m$: Threshold for negative sample \\
   \textendash\ $\lambda_{diff}$, $\lambda_{con}$: Balance parameters for the loss function\\
   \textendash\ $T$: Total number of sampling times \\
   \textendash\ $\alpha_{t}$: Parameters controlling the noise added
}

\begin{itemize}[leftmargin=*, label=$\triangleright$]
    \item Training:
\end{itemize}

\For{$t \sim \text{Uniform}(\{1, \ldots, T\})$}{

   $z_{t}^{1:N} = \text{forward}(z_0^{1:N},\epsilon^{1:N},t) \hfill \blacktriangleright  Eq.~\eqref{eq_process}$
   \renewcommand{\thefootnote}{\fnsymbol{footnote}}
  $\epsilon_{\theta'\footnotemark}^{1:N} \leftarrow d_{frozen}(z_{t}^{1:N}, t, \tau_{\theta}(y))$\\
    $\Delta\text{calculate }M_{\text{Versatile} }\hfill \blacktriangleright  Eq.~\eqref{m1}$ \\
  $\Delta\text{calculate }M_{\text{Sparse-Causal}} \hfill \blacktriangleright Eq.~\eqref{m2}$ \\
  $\epsilon_{\theta}^{1:N} \leftarrow f(\epsilon_{\theta'}^{1:N}, M_{\text{Versatile}}, M_{\text{Sparse-Causal}}) \hfill $\\

  $\mathcal{L}_{diff} = \text{Diffusion\_loss}(\epsilon^{1:N}, \epsilon_{\theta}^{1:N}) \hfill \blacktriangleright Eq.~\eqref{l1}$ \\
  \For{all $i \in \{1, \ldots, N\}$}{
    $\hat{\epsilon}_{\theta}^{i} = g(\epsilon_{\theta}^{i})$ \\
  }
  \For{all $i \in \{1, \ldots, N-1\}$}{
    \text{Calculate } ${l_t}^{(i)} \hfill \blacktriangleright Eq.~\eqref{cl}$ \\
  }
  $\mathcal{L}_{con} = \text{Con\_loss}({l_t}^{(1)}, {l_t}^{(2)}, \ldots, {l_t}^{(N-1)}) \hfill \blacktriangleright Eq.~\eqref{conloss}$ \\
  $\mathcal{L} =\lambda_{diff} \cdot \mathcal{L}_{diff} + \lambda_{con} \cdot \mathcal{L}_{con}$ \\
  update network $f$ and noise-adapter $g$ to minimize $\mathcal{L}$
} \label{alg_1}

\begin{itemize}[leftmargin=*, label=$\triangleright$]
  \item Inference:
\end{itemize}
\For{$t = T$ \KwTo $1$}{
\renewcommand{\thefootnote}{\fnsymbol{footnote}}
$\epsilon_{\theta'\footnotemark[\value{footnote}]}^{1:N} \leftarrow d_{frozen}(z_{t}^{1:N}, t, \tau_{\theta}(y))$ \\
  $\Delta\text{calculate }M_{\text{Versatile}}\hfill \blacktriangleright  Eq.~\eqref{m1}$  \\
  $\Delta\text{calculate }M_{\text{Sparse-Causal}} \hfill \blacktriangleright Eq.~\eqref{m2}$ \\
  $\epsilon_{\theta}^{1:N} \leftarrow f(\epsilon_{\theta'}^{1:N}, M_{\text{Versatile}}, M_{\text{Sparse-Causal}}) \hfill $\\
  $z_{t-1}^{1:N} = \text{DDIM\_sample}(z_t^{1:N}, \epsilon_{\theta}^{1:N}) \hfill \blacktriangleright  Eq.~\eqref{ddim}$
}
$z_{0}^{1:2N-1} = \text{SOFP\_Interpolation}(z_0^{1:N}) \hfill $
\end{algorithm}

\section{Experimentation}
\subsection{Experimental Setup}
All our experiments are conducted on a single NVIDIA GeForce RTX 3090 with 24GB of VRAM. We perform all computational tasks on a Linux-based system running Python 3.10.13, compiled with the GCC 11.2.0 compiler. To ensure efficient execution of parallel computing tasks, we employ the NVIDIA CUDA 12.0 platform.

\subsubsection{\textbf{Training Settings}} We directly use the trained Stable Diffusion v1.5 as the base model. We train the motion module and noise adapter using the proposed CLV-HD dataset. The video clips in the dataset are first sampled at a stride of 4, then resized and center-cropped to a resolution of $256 \times 256$. The frame length is set to 8, the batch size to 1, the learning rate to $1 \times 10^{-5}$, and the total number of training steps to 40k. In the contrastive learning of noise, we set the threshold for negative samples to 4, and the temperature parameter $\tau$ to 0.07. We use a linear beta schedule as in AnimateDiff~\cite{Guo2023AnimateDiffAY}, where $\beta_{\text{start}} = 0.00085$ and $\beta_{\text{end}} = 0.012$. For loss function, we set $\lambda_{diff}$ as 1 and $\lambda_{con}$ as 0.07.

\subsubsection{\textbf{Inference Settings}}At inference, we use DDIM~\cite{Song2020DenoisingDI} sampler in our experiments. We only use our frame interpolation module during inference. Our frame interpolation strategy expands an N-frame video to 2N-1 frames by inserting new frames between the original adjacent frames. Both the number of sampling steps and the scale for text guidance are selectable for users. When conducting qualitative and quantitative comparisons, we use DDIM with 25 sampling steps, and the scale for text guidance is 8.

\renewcommand{\thefootnote}{\fnsymbol{footnote}}
\footnotetext{$\theta'$ indicates that the results are obtained from a model with frozen parameters.}
\renewcommand{\thefootnote}{\arabic{footnote}}
\setcounter{footnote}{0}
\subsection{Results}

In this paper, we present several qualitative results from our ConCLVD model, displaying only four frames per animation due to space limitations. We encourage readers to visit our website for higher-quality visuals. Our method skillfully combines brushwork beauty with the dynamic play of light and shadow typical of Chinese landscape painting into the T2V model. It uses the ``five shades of ink''\footnote{The ``five shades of ink'' technique is a traditional Chinese painting method that uses varying depths of ink to create a spectrum of shades.} technique, evident in the second row's right-side images, where varying ink depths create nuanced color changes, beautifully transitioning across elements like mountains and water.
Additionally, our model employs the ``combined colored ink''\footnote{The ``combined colored ink'' technique refers to a method in Chinese painting where different colored inks are blended together to achieve harmonious and unified effects.} technique to blend multiple colors harmoniously, as seen in the third row, offering a synchronized and aesthetically cohesive landscape portrayal. In addition, we find our method effectively differentiates main subjects from their surroundings. For example, the animation  in the first row’s right-side showcases raindrops and a house moving at distinct speeds and blurs, creating vivid scenes. Results affirm that ConCLVD adeptly transfers Chinese landscape painting's brushwork, techniques, and style to digital media, enriching modern digital content with traditional artistic values.

\begin{figure}[!t]
    \centering
    \includegraphics[width=1\linewidth]{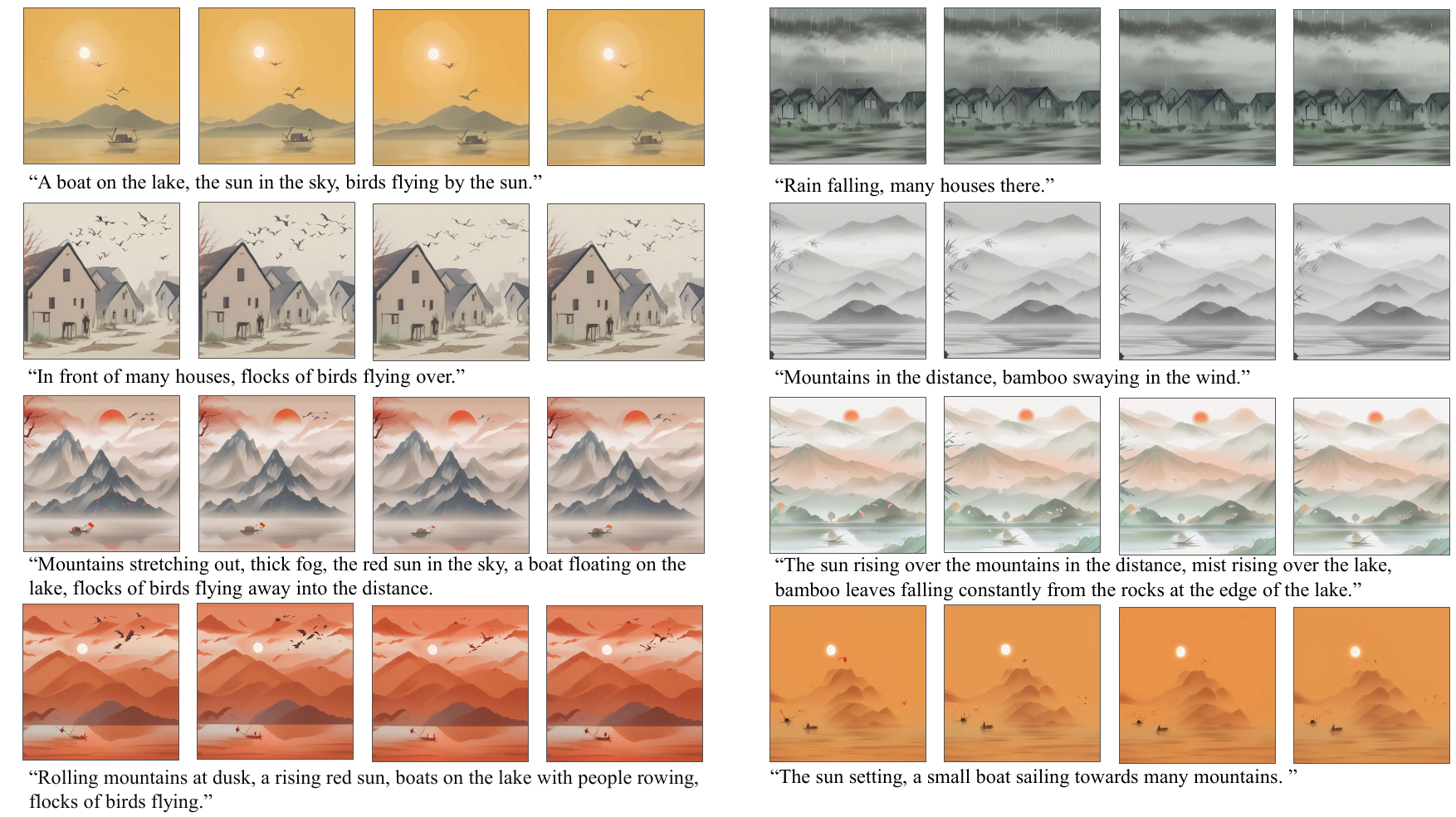}
    \captionsetup{font=small}
    \caption{Main Results. Our ConCLVD creates high-quality videos in the style of Chinese landscape painting.}\label{fig_Results show}
    \vspace{-1em}
\end{figure}

\subsection{Baseline Comparisons}

\begin{figure}[!t]\label{fig_comparionstudy}
    \centering
    \includegraphics[width=1\linewidth]{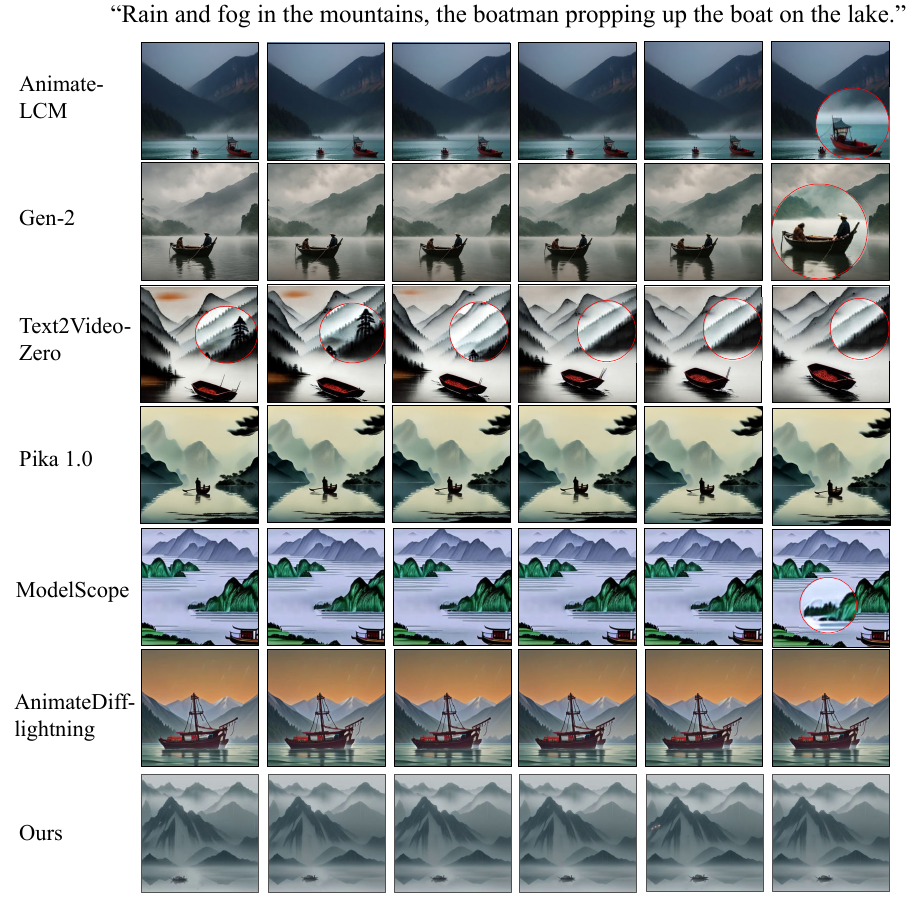}
    \captionsetup{font=small}
    \caption{Qualitative comparisons with baseline methods. Best viewed with zoom-in.}
    \label{fig_Comparisonstudy}
    \vspace{-1.5em}
\end{figure}

\begin{figure}[!t]\label{fig_fourimages}
    \includegraphics[width=1\linewidth]{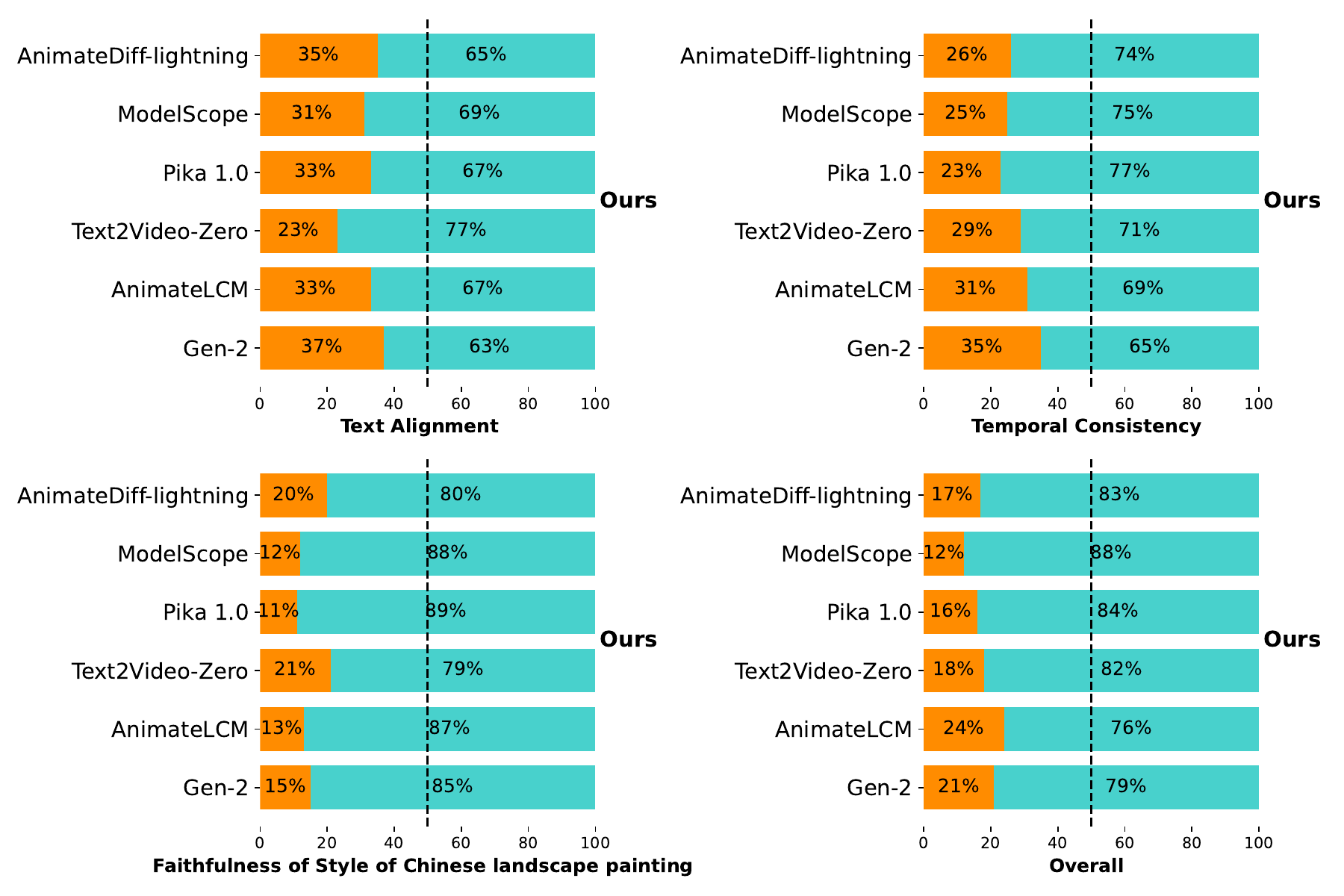}
    \captionsetup{font=small}
    \caption{User preference studies. ConCLVD outperforms all
baselines from overall aspects.}
    \label{fig_user}
\vspace{-1.5em}
\end{figure}


We evaluate our method against six prominent baselines: AnimateLCM~\cite{wang2024animatelcm}, which accelerates personalized diffusion models for high-fidelity animation via consistency decoupling; Gen-2~\cite{runwaymlGen2Runway}, a multimodal AI system for novel video creation from text; Text2Video-Zero~\cite{khachatryan2023text2videozero}, enhancing video consistency with frame-inter attention for zero-shot text-based video generation; Pika 1.0~\cite{pikaPika}, specializing in Text-to-3D special effects with parallel sampling for cinema-grade outputs; AnimateDiff-lightning~\cite{lin2024animatedifflightning}, focusing on rapid video production through progressive adversarial diffusion distillation; and ModelScope~\cite{modelscope}, which leverages a 3D-Unet architecture to iteratively denoise Gaussian noise into video content.

\subsubsection{\textbf{Qualitative Results}}

Figure \ref{fig_Comparisonstudy} presents our qualitative results, comparing our ConCLVD model with other generation models. While Gen-2~\cite{runwaymlGen2Runway} produces videos reflecting the text concept, its style deviates from Chinese landscape painting, evident in unrealistic details like the fisherman's attire. Similarly, AnimateLCM~\cite{wang2024animatelcm} and AnimateDifflightning~\cite{lin2024animatedifflightning} introduce modern elements and an overly detailed depiction, misaligning with the traditional painting style. Text2Video-Zero~\cite{khachatryan2023text2videozero} aligns closer to the landscape style but suffers from instability and inconsistency in temporal context, like fluctuating pine tree shapes. Pika 1.0~\cite{pikaPika} achieves consistency but lacks dynamic variation, and ModelScope~\cite{modelscope} fails in vivid and dynamic depiction, rendering water waves as mere lines. In contrast, our ConCLVD excels by incorporating raindrops, mist, and boats moving at varied speeds, capturing the dynamic essence of ink wash painting and harmoniously blending movement with stillness, accurately reflecting landscape painting aesthetics. Further  comparisons are available in our supplementary materials.

\subsubsection{\textbf{Quantitative Results}}


Our ConCLVD is assessed against baselines using both automatic metrics and user evaluations. For quantitative indicators, we leverage CLIP-temp~\cite{esser2023structure} for temporal consistency by computing cosine similarities between consecutive frames and CLIP-text~\cite{Qi2023FateZeroFA} for text alignment by averaging CLIP scores between video frames and prompts. We further evaluate the training computational requirements of ConCLVD with baselines.

Table~\ref{table:comparison} presents a detailed comparison of ConCLVD with baselines. Quantitative metric analysis shows that ConCLVD performs better in almost all metrics compared to baselines, albeit with slightly lower text alignment than Gen-2, which we believe is due to the frozen Stable Diffusion parameters. However, our strategy enables efficient training on an RTX 3090, balancing technical performance with training practicability.

\subsubsection{\textbf{Cost-effective with High Efficacy}} \label{ suanli}
The ``GPUs for Training'' column in Table~\ref{table:comparison} compares the training computing power requirements for ConCLVD and baselines. Pika 1.0 and Gen-2, commercial products that are not open source, are presumed to have much higher computational requirements than ConCLVD based on the estimation of their parameters. According to the article by Tim Dettmers~\cite{dettmers2023bestgpus} on selecting GPUs, given ModelScope's total parameter count of approximately 1.7 billion and a minimum hardware requirement of at least 16GB of GPU memory, it is inferred that suitable GPUs for this model is NVIDIA's RTX 3090. Text2Video-Zero is a model that requires no training, but it falls short regarding smooth inter-frame transitions. AnimateLCM's training uses eight A800 GPUs and AnimateDiff-lightning's training uses 64 A100 GPUs, whereas our ConCLVD only needs a single RTX 3090. ModelScope's training requirements are similar to ours, but the videos it generates are noticeably inferior in terms of fidelity to the landscape painting style, and the portrayal of the elements. ConCLVD allows users to train their own Chinese landscape painting video generation model with minimal cost.

\begin{table}[!t]
\caption{\textbf{Quantitative results with baselines.Text and Temp denote CLIP-text and CLIP-temp respectively. ADL stands for AnimateDiff-lightning. * represents inferred result.}}
\centering
\begin{tabular}{lccc}
\hline
Method & \multicolumn{2}{c}{Metric} & GPUs for Training \\
 & Text$\uparrow$ & Temp$\uparrow$ &   \\
\hline
Pika 1.0~\cite{pikaPika} & 24.564 & 0.972 & details not disclosed \\
T2V-Zero~\cite{khachatryan2023text2videozero} & 27.633 & 0.980 & no training required \\
AnimateLCM~\cite{wang2024animatelcm} & 29.987 & 0.970 & 8 A800 \\
Gen-2~\cite{runwaymlGen2Runway} & \textbf{30.609} & 0.978 & details not disclosed \\
ADL~\cite{lin2024animatedifflightning} & 30.421 & 0.962 & 64 A100 \\
ModelScope~\cite{modelscope} & 29.562 & 0.961 & RTX 3090* \\
\hline
\textbf{ConCLVD} & 30.518 & \textbf{0.983} & RTX3090 \\
\hline
\end{tabular}
\label{table:comparison}
\vspace{-2.0em}
\end{table}
\subsubsection{\textbf{User Study}}


100 participants, including creators, enthusiasts, and aesthetes of Chinese landscape painting, were surveyed to evaluate our generated videos for Text Alignment, Temporal Consistency, adherence to Chinese landscape painting style, and overall quality. Due to the lack of uniform quantitative indicators for  style fidelity, these participants' insights are crucial for our research. Figure~\ref{fig_user} shows our ConCLVD model significantly outshining all baseline methods in the user study. Although ConCLVD's CLIP-text score doesn't top Gen-2's, its strong stylistic fidelity to Chinese landscape painting likely influences respondents' preferences towards its text alignment. This highlights how subjective style preferences can profoundly affect perceptions, indicating that a distinctive style may have a more lasting impact.

\subsection{Ablation Studies}
\begin{figure}[!t]\label{atten}
    \centering
    \includegraphics[width=1\linewidth]{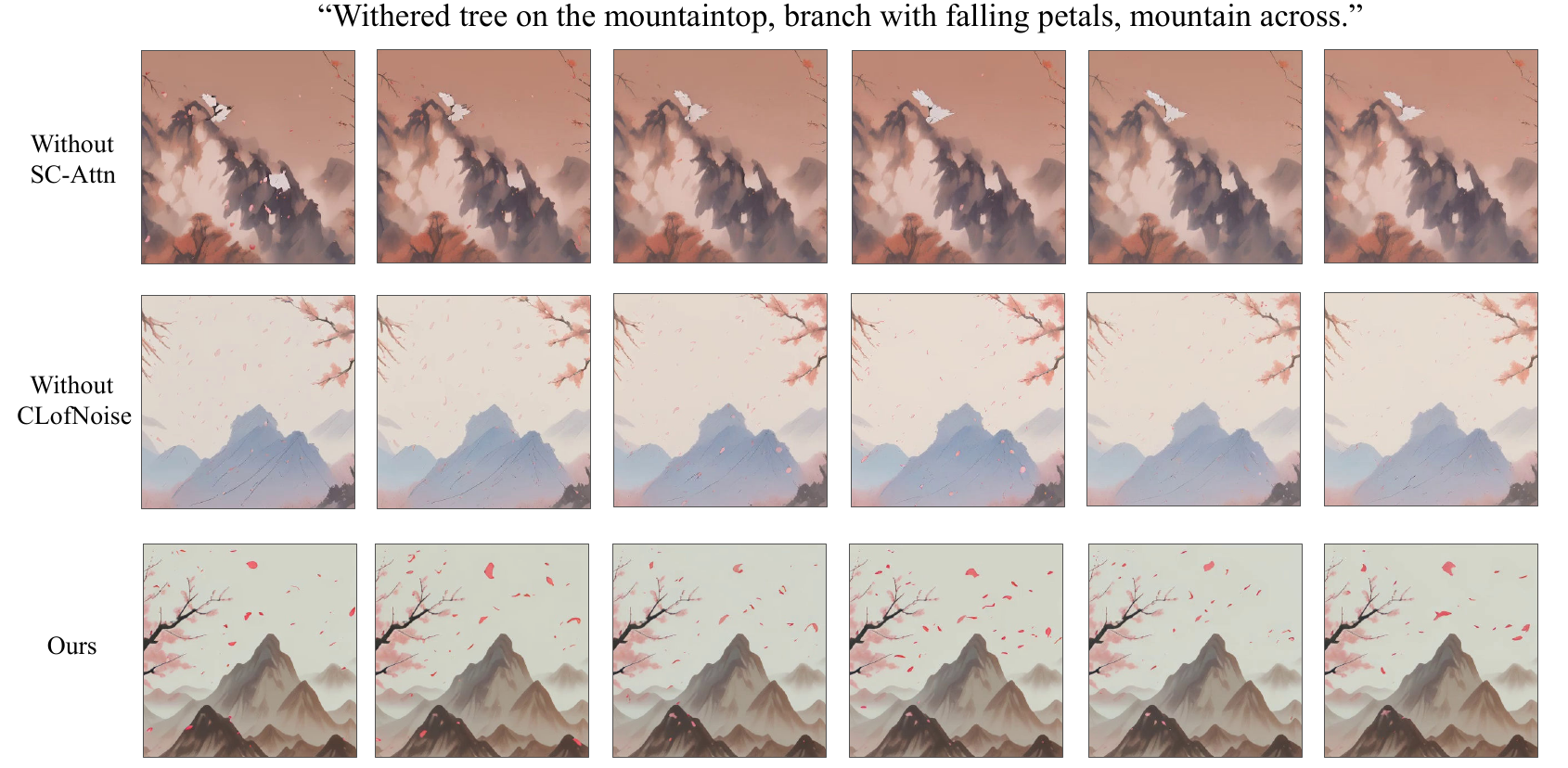}
    \captionsetup{font=small}
    \caption{Ablation study on Sparse-Casual Attention and Contrastive Learning of Noise.}
    \label{fig_Ablationstudy}
    \vspace{-0.5em}
\end{figure}

\begin{table}[!t]
\caption{Quantitative results of Ablation study.}
\centering
\begin{tabular}{lcccc}
\hline
\textbf{Model} & \textbf{Clip-Text} & \textbf{Clip-Temp}  \\
\hline
without SC-Attn & 28.862 & 0.973 \\
without CLofNoise & 29.864 & 0.964  \\
\hline
\textbf{ConCLVD} & \textbf{30.526} & \textbf{0.981} &  \\
\hline
\end{tabular}
\label{table:results}
\vspace{-1.0em}
\end{table}

We primarily conducted ablation studies on the proposed Sparse-Causal Attention (SC-Attn) within the motion module's residual architecture, and the Contrastive Learning of Noise for Video Frames (CLofNoise) , individually removing each design to assess its impact. The findings, illustrated in Figure~\ref{fig_Ablationstudy}, reveal that omitting SC-Attn leads to discontinuous frame transitions while maintaining content consistency. Conversely, excluding CLofNoise maintains content uniformity and smooth transitions but lacks in capturing dynamic changes, such as petal movement in the video.

The quantitative results, detailed in Table~\ref{table:results}, highlight the significance of both SC-Attn and CLofNoise in enhancing textual accuracy and temporal coherence. These qualitative and quantitative assessments confirm the vital roles of our model's components in achieving high-quality outputs. Supplementary materials further underscore the contributions of these key designs.

\section{Conclusion}

In this paper, we proposed a text-to-video conversion network, termed ConCLVD (Controllable Chinese Landscape Video Diffusion), focusing on generating videos of Chinese landscape painting style. Utilizing the proposed CLV-HD (Chinese Landscape Video-High Definition) dataset, motion module, contrastive learning of noise in latent space, and a frame interpolation strategy based on optical flow projection, our framework can effectively capture the dynamic beauty of landscape painting imagery and create a poetic and picturesque essence while maintaining low computational power requirements. Extensive experiments demonstrated the effectiveness of our method in visual quality, ink style prominence, and coherence, showcasing its potential in artistic video generation. We believe that our framework provides an opportunity to promote the preservation, development, and promotion of Chinese landscape painting. We also hope that our work can lead to inspiration for possible future works, bringing new energy to traditional art and culture using modern technology.

\bibliographystyle{plain}
\bibliography{sample-base}

\begin{thebibliography}{10}

\bibitem{An2023LatentShiftLD}
Jie An, Songyang Zhang, Harry Yang, Sonal Gupta, Jia-Bin Huang, Jiebo Luo, and
  Xiaoyue Yin.
\newblock Latent-shift: Latent diffusion with temporal shift for efficient
  text-to-video generation.
\newblock {\em ArXiv}, 2023.

\bibitem{ChangMM2022}
Xiang Chang, Fei Chao, Changjing Shang, and Qiang Shen.
\newblock Sundial-gan: A cascade generative adversarial networks framework for
  deciphering oracle bone inscriptions.
\newblock In {\em Proceedings of the 30th ACM International Conference on
  Multimedia}, 2022.

\bibitem{simCLR}
Ting Chen, Simon Kornblith, Mohammad Norouzi, and Geoffrey~E. Hinton.
\newblock A simple framework for contrastive learning of visual
  representations.
\newblock In {\em Proceedings of the 37th International Conference on Machine
  Learning, {ICML} 2020, 13-18 July 2020, Virtual Event}, 2020.

\bibitem{Child2019GeneratingLS}
Rewon Child, Scott Gray, Alec Radford, and Ilya Sutskever.
\newblock Generating long sequences with sparse transformers.
\newblock {\em ArXiv}, 2019.

\bibitem{Deng2023EfficiencyoptimizedVD}
Zijun Deng, Xiangteng He, and Yuxin Peng.
\newblock Efficiency-optimized video diffusion models.
\newblock {\em Proceedings of the 31st ACM International Conference on
  Multimedia}, 2023.

\bibitem{Deng2023MVDiffusionMV}
Zijun Deng, Xiangteng He, Yuxin Peng, Xiongwei Zhu, and Lele Cheng.
\newblock Mv-diffusion: Motion-aware video diffusion model.
\newblock {\em Proceedings of the 31st ACM International Conference on
  Multimedia}, 2023.

\bibitem{dettmers2023bestgpus}
Tim Dettmers.
\newblock The best gpus for deep learning in 2023 --- an in-depth analysis,
  2023.

\bibitem{dhariwal2021diffusion}
Prafulla Dhariwal and Alex Nichol.
\newblock Diffusion models beat gans on image synthesis, 2021.

\bibitem{esser2023structure}
Patrick Esser, Johnathan Chiu, Parmida Atighehchian, Jonathan Granskog, and
  Anastasis Germanidis.
\newblock Structure and content-guided video synthesis with diffusion models,
  2023.

\bibitem{modelscope}
Institute for Intelligent~Computing.
\newblock Text-to-video-synthesis model in open domain --- modelscope.cn.
\newblock
  \url{https://modelscope.cn/models/iic/text-to-video-synthesis/summary}, 2024.

\bibitem{Geyer2023TokenFlowCD}
Michal Geyer, Omer Bar-Tal, Shai Bagon, and Tali Dekel.
\newblock Tokenflow: Consistent diffusion features for consistent video
  editing.
\newblock {\em ArXiv}, 2023.

\bibitem{Guo2023AnimateDiffAY}
Yuwei Guo, Ceyuan Yang, Anyi Rao, Yaohui Wang, Y.~Qiao, Dahua Lin, and Bo~Dai.
\newblock Animatediff: Animate your personalized text-to-image diffusion models
  without specific tuning.
\newblock {\em ArXiv}, 2023.

\bibitem{Gupta2018ImagineTS}
Tanmay Gupta, Dustin Schwenk, Ali Farhadi, Derek Hoiem, and Aniruddha Kembhavi.
\newblock Imagine this! scripts to compositions to videos.
\newblock In {\em European Conference on Computer Vision}, 2018.

\bibitem{Hertz2022PrompttoPromptIE}
Amir Hertz, Ron Mokady, Jay~M. Tenenbaum, Kfir Aberman, Yael Pritch, and Daniel
  Cohen-Or.
\newblock Prompt-to-prompt image editing with cross attention control.
\newblock {\em ArXiv}, 2022.

\bibitem{Park2021AsymmetricBM}
Jun ho~Park, Chulwoo Lee, and Chang-Su Kim.
\newblock Asymmetric bilateral motion estimation for video frame interpolation.
\newblock 2021.

\bibitem{Huang2020RealTimeIF}
Zhewei Huang, Tianyuan Zhang, Wen Heng, Boxin Shi, and Shuchang Zhou.
\newblock Real-time intermediate flow estimation for video frame interpolation.
\newblock In {\em European Conference on Computer Vision}, 2020.

\bibitem{khachatryan2023text2videozero}
Levon Khachatryan, Andranik Movsisyan, Vahram Tadevosyan, Roberto Henschel,
  Zhangyang Wang, Shant Navasardyan, and Humphrey Shi.
\newblock Text2video-zero: Text-to-image diffusion models are zero-shot video
  generators.
\newblock In {\em {IEEE/CVF} International Conference on Computer Vision,
  {ICCV} 2023, Paris, France, October 1-6, 2023}, 2023.

\bibitem{LiLZWXWZP0J23}
Lijiang Li, Huixia Li, Xiawu Zheng, Jie Wu, Xuefeng Xiao, Rui Wang, Min Zheng,
  Xin Pan, Fei Chao, and Rongrong Ji.
\newblock Autodiffusion: Training-free optimization of time steps and
  architectures for automated diffusion model acceleration.
\newblock In {\em {IEEE/CVF} International Conference on Computer Vision,
  {ICCV} 2023, Paris, France, October 1-6, 2023}, 2023.

\bibitem{Lin2018TransformAS}
Daoyu Lin, Yang Wang, Guangluan Xu, Jun~Yu Li, and Kun Fu.
\newblock Transform a simple sketch to a chinese painting by a multiscale deep
  neural network.
\newblock {\em Algorithms}, 2018.

\bibitem{lin2024animatedifflightning}
Shanchuan Lin and Xiao Yang.
\newblock Animatediff-lightning: Cross-model diffusion distillation, 2024.

\bibitem{Liu2019CrossModalDL}
Yue Liu, Xin Wang, Yitian Yuan, and Wenwu Zhu.
\newblock Cross-modal dual learning for sentence-to-video generation.
\newblock {\em Proceedings of the 27th ACM International Conference on
  Multimedia}, 2019.

\bibitem{Ma2023FollowYP}
Yue Ma, Yin-Yin He, Xiaodong Cun, Xintao Wang, Ying Shan, Xiu Li, and Qifeng
  Chen.
\newblock Follow your pose: Pose-guided text-to-video generation using
  pose-free videos.
\newblock In {\em AAAI Conference on Artificial Intelligence}, 2023.

\bibitem{Marwah2017AttentiveSV}
Tanya Marwah, Gaurav Mittal, and Vineeth~N. Balasubramanian.
\newblock Attentive semantic video generation using captions.
\newblock {\em 2017 IEEE International Conference on Computer Vision (ICCV)},
  2017.

\bibitem{Mittal2016SyncDRAWAV}
Gaurav Mittal, Tanya Marwah, and Vineeth~N. Balasubramanian.
\newblock Sync-draw: Automatic video generation using deep recurrent attentive
  architectures.
\newblock {\em Proceedings of the 25th ACM International Conference on
  Multimedia}, 2016.

\bibitem{Pan2017ToCW}
Yingwei Pan, Zhaofan Qiu, Ting Yao, Houqiang Li, and Tao Mei.
\newblock To create what you tell: Generating videos from captions.
\newblock {\em Proceedings of the 25th ACM International Conference on
  Multimedia}, 2017.

\bibitem{pikaPika}
{Pika labs}.
\newblock {P}ika --- pika.art.
\newblock \url{https://pika.art/}, 2024.
\newblock [Accessed 25-03-2024].

\bibitem{Qi2023FateZeroFA}
Chenyang Qi, Xiaodong Cun, Yong Zhang, Chenyang Lei, Xintao Wang, Ying Shan,
  and Qifeng Chen.
\newblock Fatezero: Fusing attentions for zero-shot text-based video editing.
\newblock {\em 2023 IEEE/CVF International Conference on Computer Vision
  (ICCV)}, 2023.

\bibitem{CLIP}
Alec Radford, Jong~Wook Kim, Chris Hallacy, Aditya Ramesh, Gabriel Goh,
  Sandhini Agarwal, Girish Sastry, Amanda Askell, Pamela Mishkin, Jack Clark,
  Gretchen Krueger, and Ilya Sutskever.
\newblock Learning transferable visual models from natural language
  supervision.
\newblock In {\em International Conference on Machine Learning}, 2021.

\bibitem{latentdiffusionmodel}
Robin Rombach, A.~Blattmann, Dominik Lorenz, Patrick Esser, and Bj{\"o}rn
  Ommer.
\newblock High-resolution image synthesis with latent diffusion models.
\newblock pages 10674--10685, 2021.

\bibitem{unet}
Olaf Ronneberger, Philipp Fischer, and Thomas Brox.
\newblock U-net: Convolutional networks for biomedical image segmentation.
\newblock In {\em Medical Image Computing and Computer-Assisted Intervention -
  {MICCAI} 2015 - 18th International Conference Munich, Germany, October 5 - 9,
  2015, Proceedings, Part {III}}, 2015.

\bibitem{runwaymlGen2Runway}
{RunwayML Research Team}.
\newblock {G}en-2 by {R}unway --- research.runwayml.com.
\newblock \url{https://research.runwayml.com/gen2}, 2024.
\newblock [Accessed 25-03-2024].

\bibitem{Singer2022MakeAVideoTG}
Uriel Singer, Adam Polyak, Thomas Hayes, Xi~Yin, Jie An, Songyang Zhang, Qiyuan
  Hu, Harry Yang, Oron Ashual, Oran Gafni, Devi Parikh, Sonal Gupta, and Yaniv
  Taigman.
\newblock Make-a-video: Text-to-video generation without text-video data.
\newblock 2023.

\bibitem{Song2020DenoisingDI}
Jiaming Song, Chenlin Meng, and Stefano Ermon.
\newblock Denoising diffusion implicit models.
\newblock In {\em 9th International Conference on Learning Representations,
  {ICLR} 2021, Virtual Event, Austria, May 3-7, 2021}, 2021.

\bibitem{RAFT}
Zachary Teed and Jia Deng.
\newblock Raft: Recurrent all-pairs field transforms for optical flow.
\newblock In {\em European Conference on Computer Vision}, 2020.

\bibitem{Vaswani2017AttentionIA}
Ashish Vaswani, Noam~M. Shazeer, Niki Parmar, Jakob Uszkoreit, Llion Jones,
  Aidan~N. Gomez, Lukasz Kaiser, and Illia Polosukhin.
\newblock Attention is all you need.
\newblock In {\em Neural Information Processing Systems}, 2017.

\bibitem{wang2024animatelcm}
Fu-Yun Wang, Zhaoyang Huang, Xiaoyu Shi, Weikang Bian, Guanglu Song, Yu~Liu,
  and Hongsheng Li.
\newblock Animatelcm: Accelerating the animation of personalized diffusion
  models and adapters with decoupled consistency learning, 2024.

\bibitem{Wang2023VideoFactorySA}
Wenjing Wang, Huan Yang, Zixi Tuo, Huiguo He, Junchen Zhu, Jianlong Fu, and
  Jiaying Liu.
\newblock Videofactory: Swap attention in spatiotemporal diffusions for
  text-to-video generation.
\newblock {\em ArXiv}, 2023.

\bibitem{Wang2019ChinaStyleAM}
Yuan Wang, W.~Zhang, and Peng Chen.
\newblock Chinastyle: A mask-aware generative adversarial network for chinese
  traditional image translation.
\newblock {\em SIGGRAPH Asia 2019 Technical Briefs}, 2019.

\bibitem{Wang2023CCLAPCC}
Zhong~Ling Wang, Jie Zhang, Zhilong Ji, Jinfeng Bai, and S.~Shan.
\newblock Cclap: Controllable chinese landscape painting generation via latent
  diffusion model.
\newblock {\em 2023 IEEE International Conference on Multimedia and Expo
  (ICME)}, 2023.

\bibitem{Wu2023LAMPLA}
Ruiqi Wu, Liangyu Chen, Tong Yang, Chunle Guo, Chongyi Li, and Xiangyu Zhang.
\newblock Lamp: Learn a motion pattern for few-shot-based video generation.
\newblock {\em ArXiv}, 2023.

\bibitem{Xing2023MakeYourVideoCV}
Jinbo Xing, Menghan Xia, Yuxin Liu, Yuechen Zhang, Yong Zhang, Yin-Yin He,
  Hanyuan Liu, Haoxin Chen, Xiaodong Cun, Xintao Wang, Ying Shan, and Tien-Tsin
  Wong.
\newblock Make-your-video: Customized video generation using textual and
  structural guidance.
\newblock {\em IEEE Transactions on Visualization and Computer Graphics}, 2023.

\bibitem{Xue2020EndtoEndCL}
Alice Xue.
\newblock End-to-end chinese landscape painting creation using generative
  adversarial networks.
\newblock {\em 2021 IEEE Winter Conference on Applications of Computer Vision
  (WACV)}, 2020.

\bibitem{SeCo}
Ting Yao, Yiheng Zhang, Zhaofan Qiu, Yingwei Pan, and Tao Mei.
\newblock Seco: Exploring sequence supervision for unsupervised representation
  learning.
\newblock In {\em AAAI Conference on Artificial Intelligence}, 2020.

\bibitem{Yuan2022LearningTG}
Shaozu Yuan, Aijun Dai, Zhiling Yan, Ruixue Liu, Meng Chen, Baoyang Chen,
  Zhijie Qiu, and Xiaodong He.
\newblock Learning to generate poetic chinese landscape painting with
  calligraphy.
\newblock 2022.

\bibitem{zhong2023clearer}
Zhihang Zhong, Gurunandan Krishnan, Xiao Sun, Yu~Qiao, Sizhuo Ma, and Jian
  Wang.
\newblock Clearer frames, anytime: Resolving velocity ambiguity in video frame
  interpolation, 2023.

\bibitem{Zhou2019AnIA}
Aven~Le Zhou, Qiu-Feng Wang, Kaizhu Huang, and Cheng-Hung Lo.
\newblock An interactive and generative approach for chinese shanshui painting
  document.
\newblock {\em 2019 International Conference on Document Analysis and
  Recognition (ICDAR)}, 2019.

\bibitem{Zhou2022MagicVideoEV}
Daquan Zhou, Weimin Wang, Hanshu Yan, Weiwei Lv, Yizhe Zhu, and Jiashi Feng.
\newblock Magicvideo: Efficient video generation with latent diffusion models.
\newblock {\em ArXiv}, 2022.

\end{thebibliography}


\end{document}